# Lightweight Image Enhancement Network for Mobile Devices Using Self-Feature Extraction and Dense Modulation

Sangwook Baek[1], Yongsup Park[1], Youngo Park[1], Jungmin Lee[1], and Kwangpyo Choi[1]

Convolutional neural network (CNN) based image enhancement methods such as super-resolution and detail enhancement have achieved remarkable performances. However, amounts of operations including convolution and parameters within the networks cost high computing power and need huge memory resource, which limits the applications with on-device requirements. Lightweight image enhancement network should restore details, texture, and structural information from low-resolution input images while keeping their fidelity. To address these issues, a lightweight image enhancement network is proposed. The proposed network include self-feature extraction module which produces modulation parameters from low-quality image itself, and provides them to modulate the features in the network. Also, dense modulation block is proposed for unit block of the proposed network, which uses dense connections of concatenated features applied in modulation layers. Experimental results demonstrate better performance over existing approaches in terms of both quantitative and qualitative evaluations.

## 1. INTRODUCTION

Recently, as the use of media services including Social Networking Service (SNS) has been increased, customers who transmit and share their images and videos through the online platforms by using mobile phones also have increased. Since these images are processed with downsampling and compression through the online platform server, their visual quality is noticeably degraded when stored from the server. In addition, the cases tend to increase that users capture their old-photographs by using mobile phone cameras and store them. It causes the quality worsening, since most old-photographs have some degradation characteristics such as faded color and irregular noise patterns, etc.

Therefore, there are increasing needs for image enhancement solutions, which reconstruct a high-quality image from a low-quality image. Such solutions are operated on the mobile device requirements, and can be applied for image quality enhancement including face renovation and old-photograph restoration. As depicted in Fig. 1, image enhancement solutions perform not only super-resolution with compression artifacts removal for low-resolution images, but detail improvement including color enhancement while maintaining the resolution.

Many studies have introduced deep convolutional network based image super-resolution methods [1-3]. It is obvious that CNN has become a disruptive technology and noticeably changed the landscape of image processing applications [4], [5]. They exhibit significant improvements for semantic fidelity and visual reality of reconstructed images.

However, these solutions have limitations to perform in lightweight models for SoC requirements while keeping identities of their networks. In spite of increasing demands of image enhancement methods which enable to operate in on-device environments, conventional lightweight networks from existing methods show low performance with undesired artifacts, such as aliasing effects.

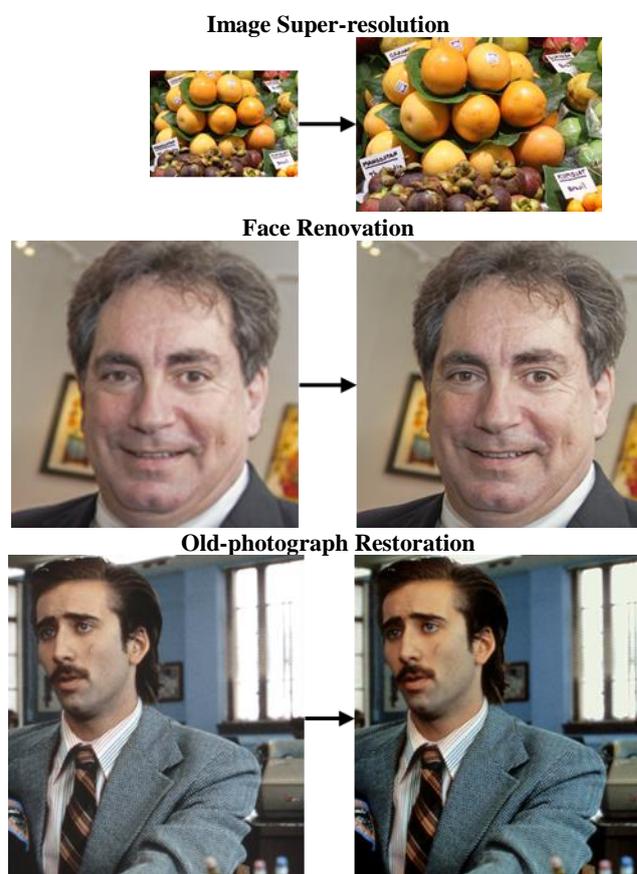

**Fig. 1. AI image enhancement solution.** It conducts two solutions: (1) image super-resolution including compression artifacts removal, (2) image enhancement including face renovation and old-photograph restoration while maintaining image resolution.

---

[1]Samsung Research, 34, Seongchon-gil, Seocho-gu, Seoul, 06765, Korea

In order to obtain high-quality images from lightweight networks, using prior information such as edge, texture, or semantic information including segmentation map can be a solution [3]. However, this information needs excessive cost to compute it, and it does not guarantee that the pre-defined information is suitable for the networks. Therefore, we propose a lightweight image enhancement network for image quality enhancement which contains super-resolution, color enhancement, face renovation, and image restoration including old-photos, etc. In the proposed network, self-feature extraction module is proposed, which aims to produce suitable information from input image itself to modulate main generator. Also, dense modulation block is proposed, which uses dense connections of concatenated features applied in modulation layers.

In this paper, we proposed two image enhancement network architectures: (1) image super-resolution (×4 upsampling factor) model including denoising and compression artifacts removal, and (2) image quality enhancement model which performs face renovation and old-photo restoration. In details, old-photo restoration model conducts not only detail improvement but color enhancement and noise reduction of recaptured old-photo images. It shows superior performance over existing methods for image quality enhancement in terms of both semantic fidelity and visual reality.

## 2. LIGHTWEIGHT IMAGE ENHANCEMENT NETWORK

### 2.1. Network architectures

As previously mentioned, the proposed networks have two architectures. Both architectures commonly consist of main generator including six dense modulation blocks and AI self-feature extraction module that generates AI features. Fig. 2. shows two proposed network architectures which can be implemented in on-device design. Basically, the proposed networks consist of convolution layers with $3 \times 3$ kernels and 16 channels, and leaky-ReLU ($\alpha = 0.1$) [6] layers are used as activation. For the image super-resolution architecture, upsampling layer composed of pixel-shuffling [7] is designed after processing the features in main generator to improve computational efficiency. In case of image enhancement architecture, the network first applies downsampling to the features by using space-to-depth transformation [8], then the features are processed by main generator, since it makes possible to refer to a wide range of image patches with small kernel size. After processing the features in main generator, upsampling layer is employed, which is the same as the image super-resolution model. Referring to larger image patches in training, it enables to process a large image context, so that the training efficiency can be increased. In addition, to provide information in the feature of each resolution before applying downsampling, the connection consisting Spatially-Adaptive Denormalization (SPADE) layer [9] is designed as a hierarchical guidance, which is explained in detail in the paragraphs 2.4.

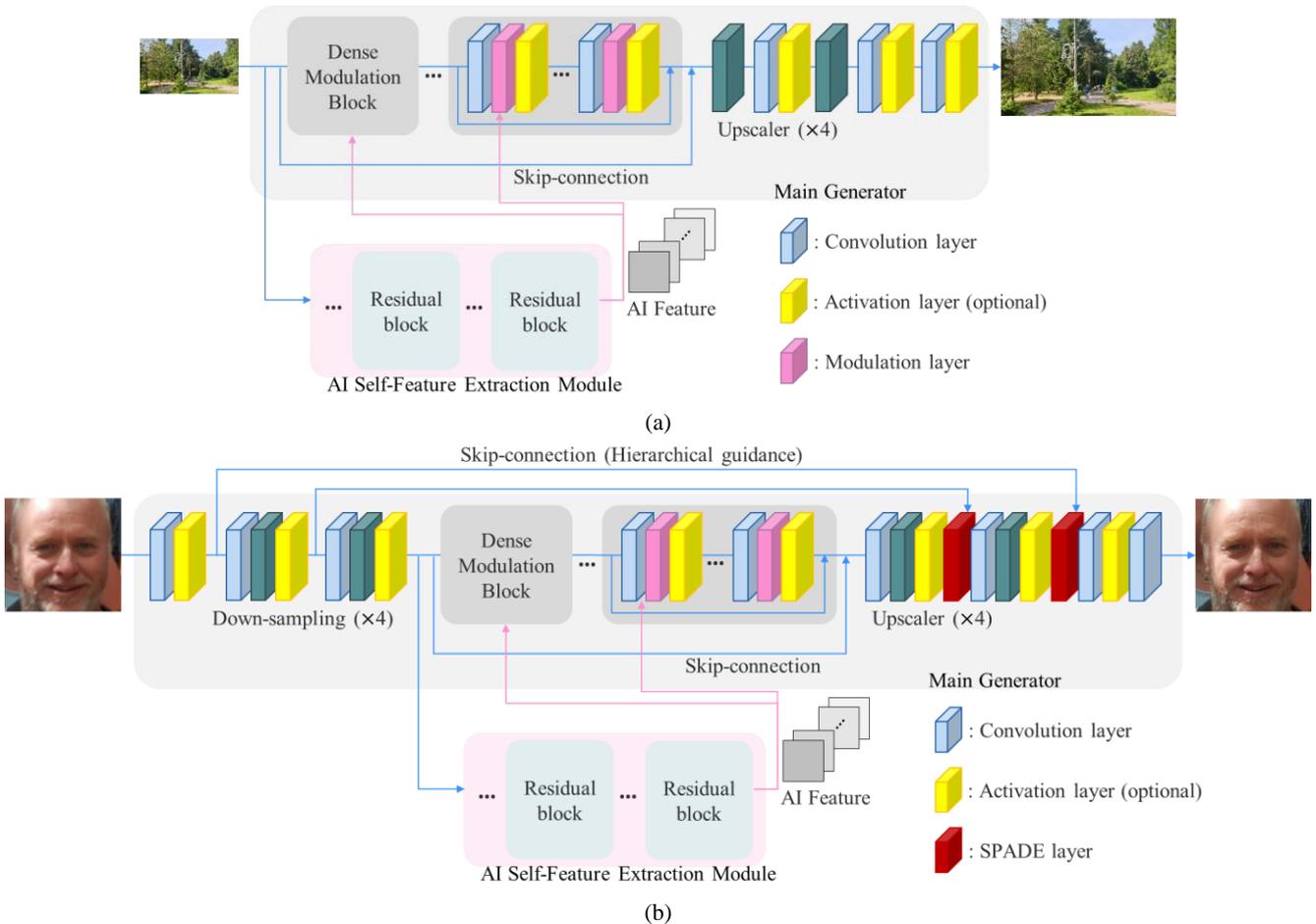

**Fig. 2. Network architectures.** (a) Super-resolution (×4) (b) Detail enhancement. The proposed network is a lightweight CNN network for reconstructing low-quality images. Overall visual quality of reconstruction is improved in both natural image and human face image. In addition, the proposed network applies color enhancement and noise reduction for restoring old-photographs.

2.2. AI self-feature extraction module

The proposed networks have AI self-feature extraction module which includes two residual blocks with four additional convolution layers. This module is simultaneously trained in an end-to-end manner with main generator, and generate the AI feature map which is essential to improve visual quality.

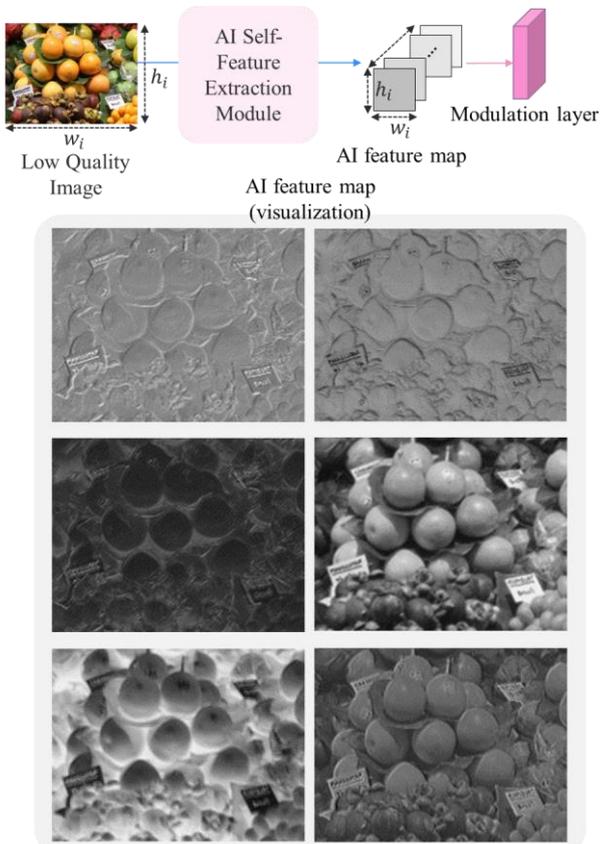

**Fig. 3. Example of AI feature map (visualization).** It contains various information such as edge, texture, and luminance, etc. AI feature maps optimize the networks by emphasizing and alleviating the feature maps from the main generator.

It is observed that when visualizing AI feature maps, it generates various features including edge, texture, transform noise and luminance information, etc., which are hard to be generated by human insights. (see Fig. 3.). By using these AI feature maps, each modulation layer produces modulation parameters which contain spatial information including edge texture, and structural information of input images. The modulation parameters decide whether to emphasize or alleviate the features in main generator. As a result, AI self-feature extraction module provides these parameters to main generator to modulate the features in modulation layers. In details, modulation layers cost a small amount of operations consisting of two convolution layers with $1 \times 1$ kernels and 16 channels. Feature modulation method results in efficient training with a small amount of operation compared to convolution. Modulating features in main generator using the parameters is defined as (1):

$$y_i = \alpha_i \cdot x_i + \beta_i \qquad (1)$$

where $\alpha_i$ and $\beta_i$ denote learned modulation parameters, and $x_i$ denotes feature map in $i$ th modulation layer.

2.3. Dense modulation block

In main generator, we propose a unit block named dense modulation block such that the features from convolution layers are channel-concatenated by dense connections, and applied to the modulation layers. Fig. 4. shows the proposed unit block architecture which it performs pixel-wise modulation with AI feature map from self-feature extraction module. Unlike existing connection methods such as skip-connection [10], the proposed approach enables to train the network while keeping information of each feature. It enables to easily learn texture detail and structural information of high-quality images without any loss from channel reduction while using only a small number of parameters. Since conventional approaches using dense connection are optimized for reconstructing structural information [2], they cause undesired artifacts such as aliasing for low-quality images including compression artifacts while reconstructing. In addition, they require large amount of operation and memory so that their performance becomes worsening in lightweight models. However, dense modulation block generates structural information while maintaining natural details trained from the high-quality images. Also, it is observed that the proposed unit block efficiently learns the degradations in old-photographs such as saturated color and noise pattern.

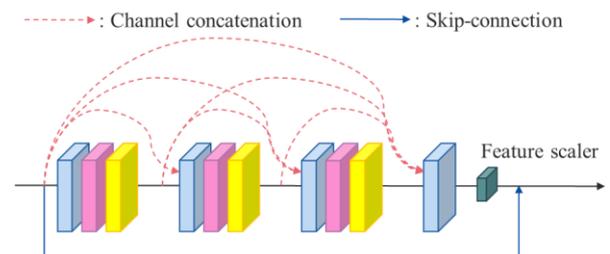

**Fig. 4. Dense modulation block architecture.** The features from convolution layers are channel-concatenated by dense connections, and it performs pixel-wise modulation with AI feature map from self-feature extraction module.

## 2.4. SPADE layer (image detail enhancement architecture)

In image detail enhancement model, to expand receptive field efficiently, downsampled features are processed in main generator. In this process, connections are designed through SPADE layer to prevent loss of information included in the high-resolution features. Similar to modulation layer, it generates parameters from high-resolution feature, and then modulate the features in main generator with them. The difference is that a convolution layer is used in common, and an additional convolution layer is processed to make each modulation parameter, since it costs number of operations to generate the parameters to modulate high-resolution features. The modulation process in SPADE layers is same as the process in modulation layers except normalization. Using normalization layer, the training procedures can be more stable, and efficiently use information in the entire image patches [11], which is defined as follows:

$$y_i = \alpha_i \cdot \frac{x_i - \mu_i}{\sigma_i} + \beta_i \qquad (2)$$

$$\mu_i = \frac{1}{N}\sum_{j \in N} x_{i,j} \qquad (3)$$

$$\sigma_i = \sqrt{\frac{1}{N}\sum_{j \in N}[(x_{i,j})^2 - \mu_i^2]} \qquad (4)$$

where $\mu_i$ and $\sigma_i$ denote the mean and standard deviation of the feature map in $i$ th channel.

## 2.5. Training details

For training, Google Open Image Dataset [12] and Datatang Image Dataset [13] were used, which have licenses basically available for commercial use. Moreover, personal information images including human face or name tags were excluded from training to respond to domestic law as well. For training image super-resolution model, image pairs were generated by applying bicubic downsampling (×4) and JPEG encoding (q=20), which is widely used degradation intensity in the online platforms including social media services.

In case of image enhancement model, training image pairs were degraded by applying bicubic downsampling and then applying upsampling (×4), which can be used in many applications such as face renovation and detailed texture improvement in low-quality images.

However, since its corresponding high-quality image for old-photograph does not exist, it is hard to generate training datasets from real images. Therefore, we collected recaptured images of old-photographs from non-experts, and then observed the degradation characteristics from the images. Such observation was used to simulate degradation to training datasets. It shows visual quality degradations such as faded color, color saturation, and depth noise pattern, etc. Fig. 5. shows an example of training datasets for old-photograph restoration model.

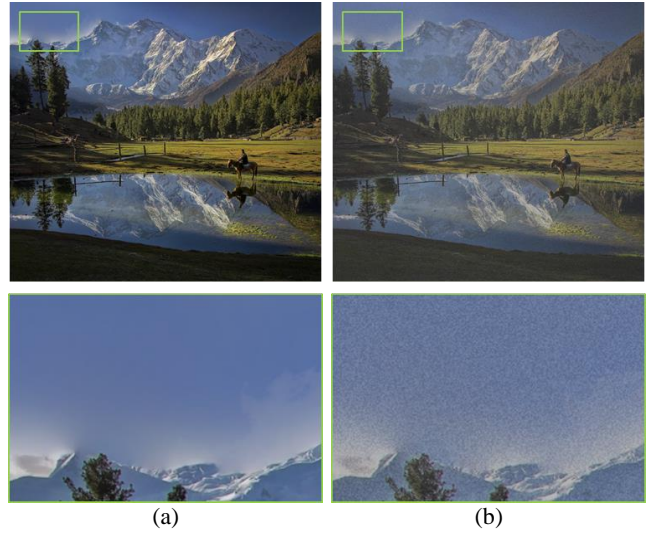

(a)          (b)

**Fig. 5. Example of training datasets for old-photograph restoration model.** (a) original image (b) degraded image. In order to restore visual quality degradation in old-photographs, training datasets were generated by applying faded color, color saturation, and noise from the observed range of collected old-photographs.

In order to optimize the network, several efficient loss functions were applied. To optimize image super-resolution model, mean square error (MSE) loss (5) and structural similarity (SSIM) loss (6) which aims to restore structural information from high-resolution (HR) images were used.

$$L^{\text{MSE}}(x, y) = \frac{1}{N}\sum_{i \in N} | y_i - f(x_i) | \qquad (5)$$

$$L^{\text{SSIM}}(x, y) = 1 - \frac{(2\mu_{f(x)}\mu_y + c_1)(2\sigma_{f(x)y} + c_2)}{(\mu_{f(x)}^2 + \mu_y^2 + c_1)(\sigma_{f(x)}^2 + \sigma_y^2 + c_2)} \qquad (6)$$

where $x$ and $y$ denote input and target image, and $f(x)$ denotes reconstructed image.

In addition to the loss functions mentioned above, VGG19 based perceptual loss [14] and adversarial loss were applied in image enhancement model, which enables to improve both reliability and fidelity in reconstructed images, represented as follows:

$$L_j^{\text{Perceptual}}(x, y) = \frac{1}{N}\sum_{i \in N} | \phi_j(y_i) - \phi_j(f(x_i)) | \qquad (7)$$

where $\phi_j$ denotes the outputs of the $j$ th layer of VGG19 network.

For adversarial training, the relativistic GAN [15] based discriminator was employed. Therefore, the proposed networks benefit from computing gradients from both generated images and real images, while in standard GAN only generated part takes effect, which can be denoted as follows:

$$L_G^{\text{Relative GAN}}(x, y) = -\mathbf{E}_x[\log(1 - D(x, f(x)))]$$
$$-\mathbf{E}_{f(x)}[\log(D(f(x), x))], \quad (8)$$
$$D(x, y) = C(x) - \mathbf{E}_{f(x)}[C(f(x))]$$

where $\mathbf{E}$ represents the operation of taking average for all data in the mini-batch, and $C(x)$ is the non-transformed discriminator output.

In details, each network was trained with an Adam optimizer [16]. For initial weights of the networks, the method described in [17] was used, which is a recommended method for neural networks using ReLU. Also, all the experiments were conducted over $10^6$ iterations. The initial learning rate of the proposed networks was commonly set to $10^{-4}$, and halved at every $10^5$ iterations, which was decreased 4 times in total.

## 3. EXPERIMENTAL RESULTS

### 3.1. Image super-resolution model

We compare the proposed solution with conventional super-resolution algorithms at the same complexity of real-time operation in on-device environments. For a fair comparison, the compared lightweight models are optimized using training procedures that are suitable for the lightweight model rather than the existing training procedures.

The proposed network was evaluated using 100 bicubic downsampled and encoded images (JPEG q=20) from DIV2K validation set [18], which have been widely used for validating image super-resolution methods. Also, it is general level of degradation used in the online platforms including social media services. Table 1 shows that the proposed network outperforms for both PSNR and SSIM, which represent less distortion when reconstructing high resolution images.

**Table 1. Performance Comparison (Image Super-resolution)**

|  | PSNR (dB) | SSIM |
|---|---|---|
| Bicubic | 29.32 | 0.8422 |
| SRGAN [1] | 30.51 | 0.8624 |
| ESRGAN [2] | 30.56 | 0.8627 |
| SFTGAN [3] | 30.10 | 0.8560 |
| Ours | **30.63** | **0.8744** |

The proposed solution shows better performance with minimizing the reconstruction error.

In order to perform quantitative evaluation of visual quality, VMAF [19] which has been widely used as the best visual quality assessment index is measured. Fig. 6 represents the performance comparison in terms of VMAF score and network complexity. Compared to the other methods, the proposed method shows nearly doubled performance from the baseline (bicubic) in terms of VMAF score. In details, the proposed network has shown improvement in subjective visual quality, whereas it costs less computing power (the number of parameters) than other lightweight models of the conventional image super-resolution networks. While removing compression artifacts, it effectively restored the loss of detail including the structural information in the reconstructed images. Fig. 7. shows the visual comparison of the proposed solution with state-of-the-art image super-resolution networks.

Especially, it was confirmed that the conventional method in [3] which applies the semantic information shows low performance under the condition that the information was not provided. However, the proposed solution enables to restore high quality images since it produces optimized information from AI feature extraction module, which is essential to improve visual quality.

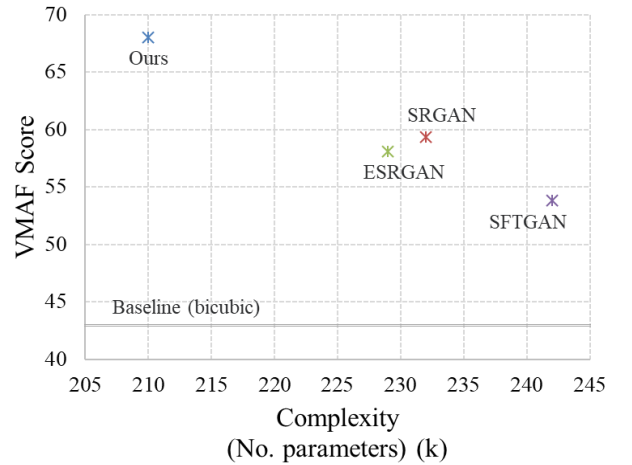

**Fig. 6. Performance comparison in terms of VMAF score and network complexity (located in upper left is better).** Proposed solution outperformed the lightweight models of conventional image super-resolution networks, whereas it costs less computation.

### 3.2. Image detail enhancement model

There have been many researches that introduce CNN based image enhancement networks including face renovation and old-photograph restoration [20-22]. However, most algorithms limit the applications with on-device environment while maintaining their identities in the networks. Therefore, it is hard to compare the proposed solutions with existing CNN algorithm. The proposed solutions were evaluated with subjective visual quality.

#### 3.2.1. Face renovation

In order to validate face renovation model, the proposed network was evaluated using bicubic downsampled (×4) and upsampled images from FFHQ validation set [23], which contains various sizes of human faces so that it has been widely used for validating face renovation methods.

Fig. 8. shows that the proposed network outperforms when reconstructing low quality images in terms of detail enhancement in human faces while keeping fidelity whereas it costs small amounts of computing power (about 300k parameters).

### 3.2.2. Old-photograph restoration

Since its ground truth image does not exist, most studies for old-photograph restoration apply their own degradation to high quality images, and evaluate quantitative performance with the restored images. In this paper, the real old-photographs captured from non-experts were collected, and we restored them by carrying out the proposed solution. Subjective visual quality was evaluated for items such as naturalness, color enhancement, and noise removal.

Fig. 9. shows that the proposed network outperforms when reconstructing low quality images in terms of old-photograph restoration. It can be observed that the degradations in old-photographs such as faded color, saturation, etc., were visually improved. Also the noise caused by the print paper pattern was effectively removed. However, it still requires a variety of technologies that cannot be handled by the proposed enhancement solutions: the solution for reflection removal appearing when recapturing the images, and the in-painting solution for damaged areas within the images. further researches on these algorithms seem to be necessary in future works.

## 4. CONCLUSION

In this paper, we proposed a lightweight network for single image quality enhancement, which enables to reconstruct a desirable high-quality image from a low-quality image. Since the proposed network is lightweight enough to be operated on a mobile device such as smartphone or tablet PC, any low-quality images from the Internet including social networks can be reconstructed in the device. In details, the proposed network performs image detail enhancement including face renovation and old-photograph restoration, which have not been lightweight for on-device requirements before. Compared with the conventional CNN based image enhancement networks, the proposed network shows improved performance in terms of detail enhancement and structural reconstruction.

## REFERENCES

[1] Ledig, Christian, et al. Photo-realistic single image super-resolution using a generative adversarial network. *In Proceedings of the IEEE conference on computer vision and pattern recognition*, 4681-4690 (2017)
[2] Wang, Xintao, et al. ESRGAN: Enhanced super-resolution generative adversarial networks. *In Proceedings of the European Conference on Computer Vision (ECCV) Workshops* (2018)
[3] Wang, Xintao, et al. Recovering realistic texture in image super-resolution by deep spatial feature transform. *In Proceedings of the IEEE conference on computer vision and pattern recognition*, 606-615 (2018)
[4] Krizhevsky, Alex, et al. Imagenet classification with deep convolutional neural networks. *Advances in neural information processing systems* **25**, 1097-1105 (2012)
[5] Szegedy, Christian, et al. Going deeper with convolutions. *In Proceedings of the IEEE conference on computer vision and pattern recognition,* 1-9 (2015)
[6] Maas, Andrew L., et al. Rectifier nonlinearities improve neural network acoustic models. *In ICML*, **30**, 3-8 (2013)
[7] Shi, Wenzhe, et al. Real-time single image and video super-resolution using an efficient sub-pixel convolutional neural network. *In Proceedings of the IEEE conference on computer vision and pattern recognition*, 1874-1883 (2016)
[8] Wang, Longguang, et al. Learning for video super-resolution through HR optical flow estimation. *In Asian Conference on Computer Vision*, 2337-2346 (2019)
[9] Park, Taesung, et al. Semantic image synthesis with spatially-adaptive normalization. *In Proceedings of the IEEE conference on computer vision and pattern recognition*, 2337-2346 (2019)
[10] Kim, Jiwon, et al. Accurate image super-resolution using very deep convolutional networks. *In Proceedings of the IEEE conference on computer vision and pattern recognition*, 1646-1654 (2016)
[11] Dumoulin, Vincent, et al. A learned representation for artistic style. *arXiv preprint arXiv:1610.07629* (2016).
[12] Google Open Image Dataset, https://storage.googleapis.com/openimages/web/index.html
[13] Datatang Image Dataset, https://www.datatang.ai/dataset/image
[14] Johnson, Justin, et al. Perceptual losses for real-time style transfer and super-resolution. *In Proceedings of the European Conference on Computer Vision (ECCV)*, 694-711 (2016)
[15] Jolicoeur-Martineau, Alexia. The relativistic discriminator: a key element missing from standard GAN. *arXiv preprint arXiv:1807.00734* (2018)
[16] Kingma, Diederik P., et al. Adam: A method for stochastic optimization. *arXiv preprint arXiv:1412.6980* (2014)
[17] He, Kaiming, et al. Delving deep into rectifiers: Surpassing human level performance on imagenet classification. *In Proceedings of the IEEE conference on computer vision*, 1026-1034 (2015)
[18] Timofte, Radu, et al. NTIRE 2018 challenge on single image super-resolution: Methods and results. *In Proceedings of the IEEE conference on computer vision and pattern recognition workshops*, (2018)
[19] Li, Zhi, et al. Toward a practical perceptual video quality metric. *The Netflix Tech Blog* **6** (2016)
[20] Chen, Yu, et al. FSRNet: End-to-end learning face super-resolution with facial priors. *In Proceedings of the IEEE conference on computer vision and pattern recognition*, 2492-2501 (2018)
[21] Yang, Lingbo, et al. HiFaceGAN: Face renovation via collaborative suppression and replenishment. *In Proceedings of the 28th ACM International Conference on Multimedia*, 1551-1560 (2020)
[22] Wan, Ziyu, et al. Bringing old photos back to life. *In Proceedings of the IEEE conference on computer vision and pattern recognition*, 2747-2757 (2020)
[23] Karras, Tero, et al. A style-based generator architecture for generative adversarial networks. *In Proceedings of the IEEE conference on computer vision and pattern recognition*, 4401-4410 (2019)

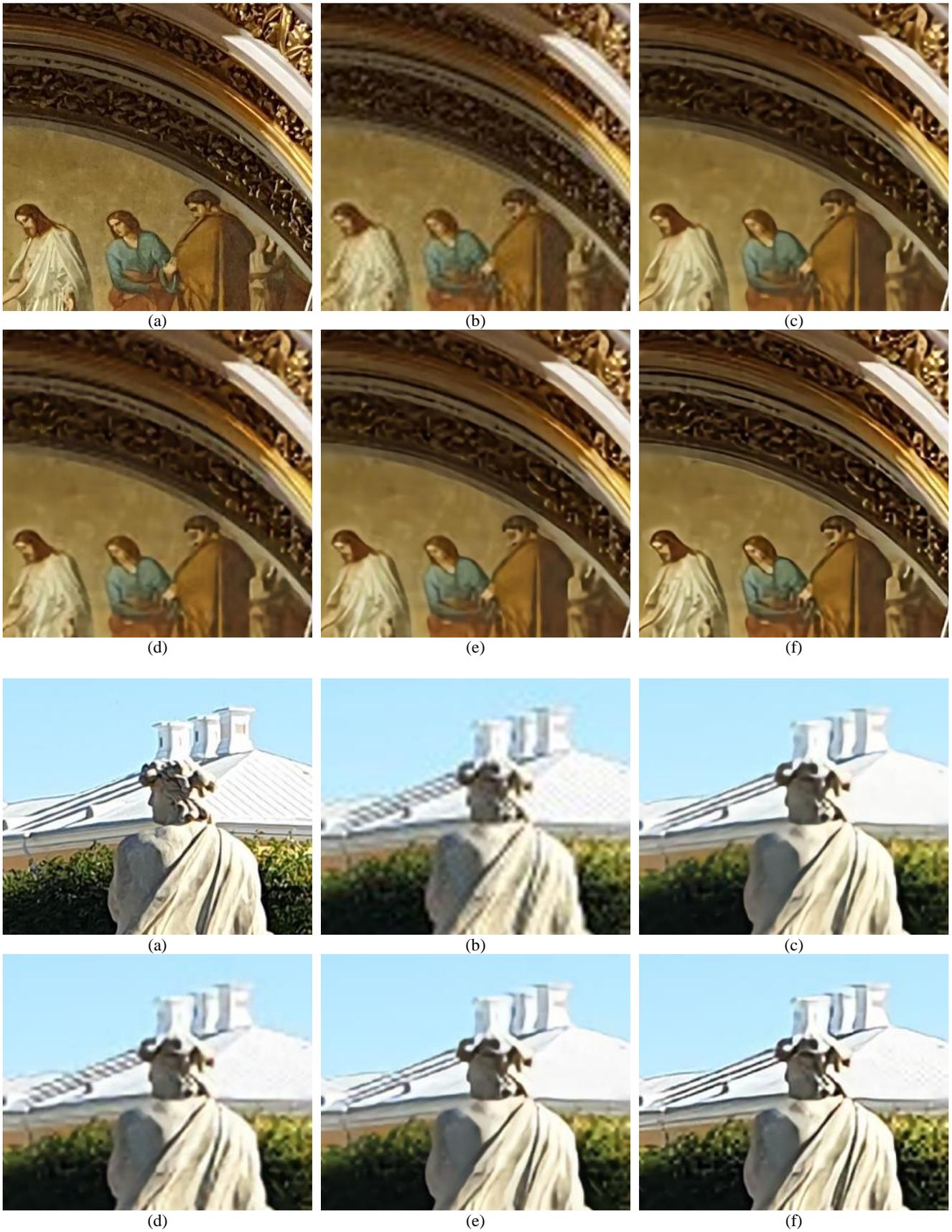

**Fig. 7. Visual quality comparison for super-resolution including compression artifacts removal (bicubic downsampling (×4) and JPEG encoding (q=20)).** (a) Ground Truth (b) Input (c) SRGAN (d) SFTGAN (without segmentation map) (e) ESRGAN (f) Ours.

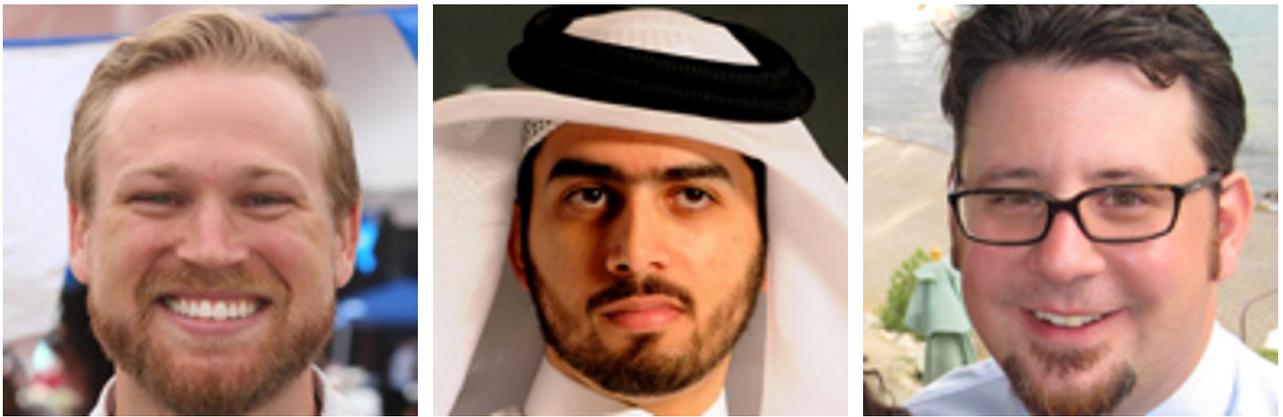
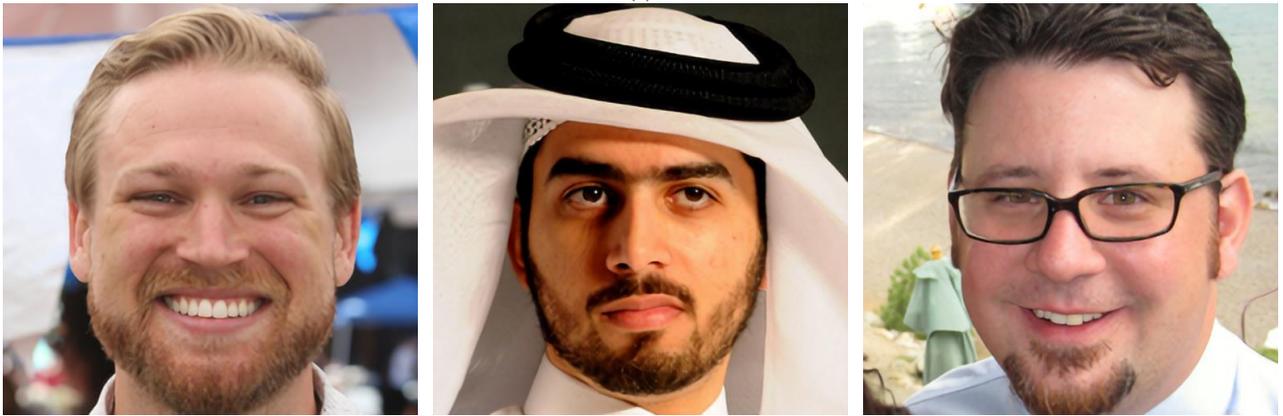

**Fig. 8. Visual quality comparison for face renovation.** (a) Input (b) Ours.

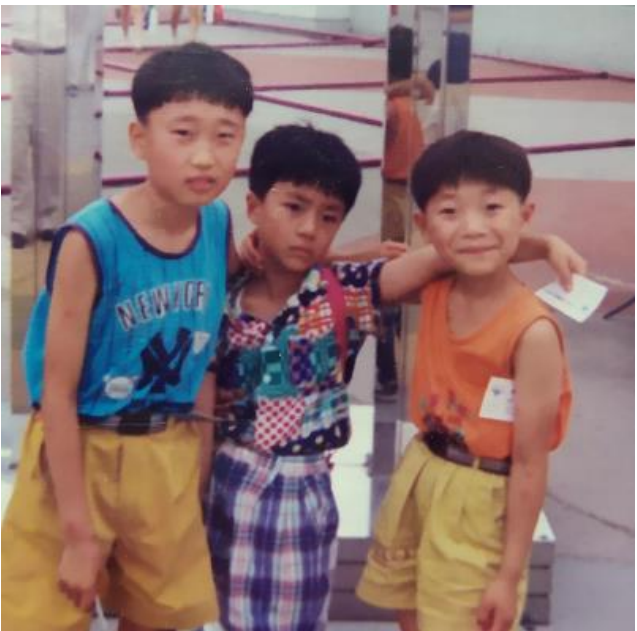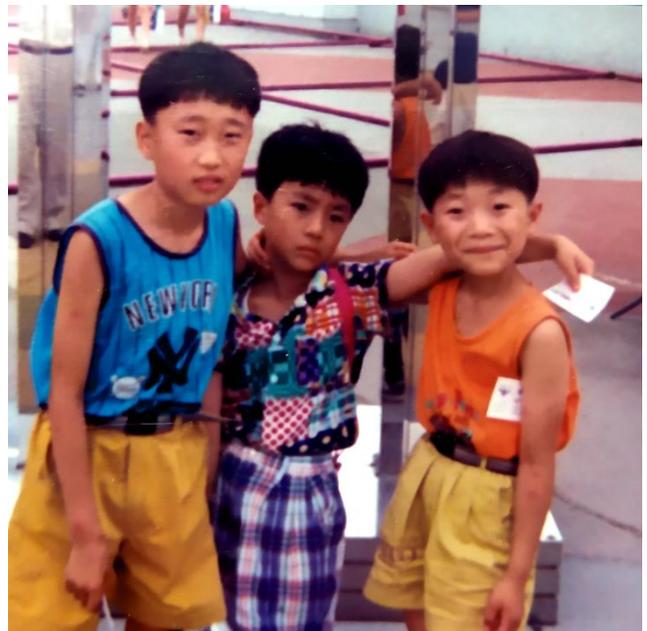

(a)          (b)

**Fig. 9. Visual quality comparison for old-photograph restoration.** (a) Input (b) Ours.